# A Deep Learning-Based Method for Metal Artifact-Resistant Syn-MP-RAGE Contrast Synthesis

Ziyi Zeng[a,b], Yuhao Wang[c], Dianlin Hu[a], T.Michael O'Shea[d], Rebecca C. Fry[d], Jing Cai[a*], Lei Zhang[b,d**],

[a] Department of Health Technology and Informatics, The Hong Kong Polytechnic University, Hong Kong SAR, China
[b] Department of Medical Physics, Duke Kunshan University, Jiangsu, China
[c] Department of Medical Physics, Duke University, Durham, North Carolina, United State
[d] University of North Carolina, Chapel Hill, North Carolina, United State

A B S T R A C T

In certain brain volumetric studies, synthetic T1-weighted magnetization-prepared rapid gradient-echo (MP-RAGE) contrast, derived from quantitative T1 MRI (T1-qMRI), proves highly valuable due to its clear white/gray matter boundaries for brain segmentation. However, generating synthetic MP-RAGE (syn-MP-RAGE) typically requires pairs of high-quality, artifact-free, multi-modality inputs, which can be challenging in retrospective studies, where missing or corrupted data is common. To overcome this limitation, our research explores the feasibility of employing a deep learning-based approach to synthesize syn-MP-RAGE contrast directly from a single channel turbo spin-echo (TSE) input, renowned for its resistance to metal artifacts. We evaluated this deep learning-based synthetic MP-RAGE (DL-Syn-MPR) on 31 non-artifact and 11 metal-artifact subjects. The segmentation results, measured by the Dice Similarity Coefficient (DSC), consistently achieved high agreement (DSC values above 0.83), indicating a strong correlation with reference segmentations, with lower input requirements. Also, no significant difference in segmentation performance was observed between the artifact and non-artifact groups.

## 1. Introduction

In brain MRI imaging studies, three-dimensional T1-weighted Magnetization Prepared—Rapid Gradient Echo (MP-RAGE) contrast is widely utilized in neuroimaging due to its clear delineation of white matter (WM) and gray matter (GM) boundaries, which is ideal for brain cortical area segmentation and volumetric analysis [1]. Open-source neuroimaging software such as FreeSurfer [2], Analysis of Functional NeuroImages (AFNI) [3], FMRIB Software Library (FSL)[4], facilitate efficient cerebral structural studies by providing automated pipelines for cerebral segmentation using atlas-based methods, and these methods require high-quality isotropic 1 mm MP-RAGE images as necessary input[5].

As an alternative for the contrast, in 2015, Nöth et al. proposed a quantitative MRI (qMRI) method to generate synthetic MP-RAGE images using SPGR/FLASH-EPI inputs, producing three types of high-contrast images: sMPR_RF_bias, sMPR_PD_T1, and sMPR_T1[6]. Among these, sMPR_T1 emerged as particularly useful in neuroimaging studies due to its unique contrast and high similarity with acquired MP-RAGE, offering outstanding visualization of the basal ganglia region and tumor boundaries for its CNR and optical contrasts performance. The resultant synthetic MP-RAGE significantly enhanced the visibility of pathological features and led to derivative clinical applications. These refinements have been applied to the volumetric study of various cerebral pathologies—including aging[7], tumors[6], [8], [9], [10], epilepsy[11], [12], [13] and focal cortical dysplasia (FCD), cortical damage in multiple sclerosis (MS) [12], [13], [14], [15], [16], spinal cord injury (SCI) [17], by facilitating morphological and volumetric analyses of regions of interest (ROI) such as white matter, gray matter cortex, basal ganglia, and cervical cord area[18], [19].

To extend the visual advantages of such contrast, Zhang et al. extended the application of artifact resistant Triple-Turbo Spin Echo (Tri-TSE) images as input for synthesizing sMPR_T1 (referred to as Syn-MPR-PD0 or Syn-PD0). By employing these TSE images with multi-parameter Bloch simulation for synthetic MP-RAGE calculation, they demonstrated that Syn-PD0 images exhibited resistance to metal artifacts while maintaining a high correlation in contrast with acq-MPR images with absolute value of correlation factor of $0.92, 0.87, 0.93$ in cerebral cortex, cerebral white matter, and subcortical gray matter segmentation DSC score respectively, between the Syn-PD0 and acquired MP-RAGE[20].

However, these mathematical based synthesis approaches require multiple designated input images acquired from specific qMRI scanning protocols with high-quality, artifact-free, precisely registered data to achieve accurate T1 maps[6], [21]. Also, as marked by Nöth, for the synthetic contrast, duration of the intermediate qMRI step is longer than that of classical anatomical imaging for comparable CNR and SNR level[6]. These situations present challenges and cost extra scanning expenditure, especially in retrospective studies, where missing or corrupted data are common.

Recent advancements in deep learning techniques offer potential solutions. Contemporary deep learning techniques in MR imaging synthesis—including Generative Adversarial Networks (GANs), conditional GANs (cGANs), and unsupervised cycleGANs—have shown great promise in mapping source modalities to target modalities through image translation tasks with less strict input requirement of using single channel [22], [23], [24]. Frameworks like Pix2Pix [25] utilizes UNET [26] architectures as generators within the GAN framework. Building upon these foundations, various specialized GAN architectures have been developed for MR contrast mapping. The edge-aware GAN (EaGAN)

* Correspondence to: J. Cai, Department of Health Technology and Informatics, The Hong Kong Polytechnic University, Hong Kong SAR, China.
** Correspondence to: L. Zhang, Department of Medical Physics, Duke Kunshan University, Jiangsu, China

*E-mail addresses:* jing.cai@polyu.edu.hk (J. Cai), lei.zhang@dukekunshan.edu.cn (L. Zhang)

incorporates Sobel edge detectors to preserve structural edges and fine details [27]. Trans-cGAN integrates transformer architectures into the GAN framework to capture long-range dependencies thus maintaining the image quality[28]. MedGAN introduces a multi-layer U-shaped architecture known as CasNet and augments the training with perceptual loss and style transfer loss functions[29]. Some other research aimed to synthesize acquired MP-RAGE contrast. Other similar DL-based works related to MP-RAGE primarily focus on synthesizing acquired MP-RAGE images, which still require the availability of scanned MP-RAGE images. For example, Ryu et al. synthesized acquired MP-RAGE images from multi-echo Gradient Recalled Echo (mGRE) sequences using a 3D U-Net architecture [30]. Similarly, Bian et al. generated 7 Tesla (7T) MP-RAGE images from 3 Tesla (3T) MP-RAGE inputs using a synthetic Generative Adversarial Network (synGAN)[31, p. 3]. Other approaches, such as the open-source SynthSR proposed by Iglesias et al. and integrated into FreeSurfer, face additional challenges as they require segmentation maps during training[5], [32]. So, there's a need to develop a deep learning implementation for synthesizing Syn-MPR with less input requirement.

For this study, we proposed a deep-learning based method for generating sMPR_T1/Syn-MPR-PD0 contrast from single scan TSE images, which we denote as DL-Syn-MPR. We design a synthesis model correspondingly. To prove the superiority of our model, we tested the preliminary synthetic ability on brain axial slices compared with other commonly used models using image quality metrics. Also, we further compared the cerebral segmentation result of our proposed synthetic contrast DL-Syn-MPR on both metal artifact subjects and normal subjects. Two special cases of other artifact subjects were tested to show clinical usability afterwards.

## 2. Methods and materials

### 2.1. Subjects

The study was approved by the Institutional Review Board (IRB) of the University of North Carolina at Chapel Hill (UNC-CH), which is one of the participants of the study. Data were collected from Extremely Low Gestational Age Newborns-Environmental influences on Child Health Outcomes (ELGAN-ECHO) study[33]. A total of 95 participants from ELGAN-ECHO study were included for our research, 41 females and 54 males, mean age $15.5 \pm 0.3$, all of whom underwent scanning using the Tri-TSE with qMRI data acquisition protocol. Among these, 51 subjects without any artifacts from the one center were selected for training.

To test the deep learning network response to the untrained imaging feature, 44 subjects from another medical centre were tested. Among these subjects, 11 subjects have metal artifacts of tooth orthodontic braces. For discussion purpose only, 1 subject have misaligned registration, 1 subject have motion artifact in one TSE acquisition. Rest 31 subjects don't have any artifacts.

### 2.2. Imaging data collection

The data collection workflow is shown on Fig.1. For the TSE acquisition, the ELGAN-ECHO protocol implemented two concatenated scans with identical scan geometry and receiver settings: a dual echo TSE (DE-TSE) and a single echo TSE (SE-TSE), collectively referred to as a Triple Turbo Spin Echo (Tri-TSE). The Tri-TSE protocol includes three weighted acquisitions: $DA_1$ = PD-weighted, $DA_2$ = T2-weighted, and $DA_3$ = T1-weighted. Imaging parameters for the Siemens Prisma 3T scanner (Siemens Medical Solutions, Erlangen, Germany) collection were as follows: voxel size of $0.5 \times 0.5 \times 2$ mm³, for qMRI calculation reason, the voxel were interpolated and resampled to $1 \times 1 \times 1$ mm³, effective echo times $TE_{\text{eff1,2}}$ of 12 ms and 102 ms, a long repetition time ($TR_{\text{long}}$) of 10 s, and a short repetition time ($TR_{\text{short}}$) of 0.5 s, Echo spacing $ES$ of 10.2 ms, with a total scan time of approximately 7: 34 minutes. Images acquired from DE-TSE sequence were registered to SE-TSE.

For the acquired MP-RAGE acquisition, the sequence follows the blocks of $[180° - \tau_1 -$ acquisition $- \tau_2]$. The acquisition parameters were $TR$ = 2400 ms, $TE$ = 2.15 ms, $TI$ = 1000 ms, flip angle $\alpha = 8°$, bandwidth = 220 Hz/Px, field of view (FoV) = $256 \times 256$ mm², resolution = $1 \times 1 \times 1$ mm³, collected in sagittal view, and accelerated with a GRAPPA PE factor of 2.

Intermediate qMRI calculation procedure follows these equations for Bloch simulation:

$$T_2 = \frac{TE_{2\text{eff}} - TE_{1\text{eff}}}{\ln\left(\frac{DA_1}{DA_2}\right)} \quad (1)$$

$$T_1 = \frac{TR_{\text{short}}}{\ln\left(\frac{DA_3}{1-DA_1}\right)} \quad (2)$$

$$PD = DA_1 e^{\left(\frac{-TE_{1\text{eff}}}{T_2}\right)} \quad (3)$$

For calculation of sMPR_T1 (Syn-PD0), $N_3D$=160 slices, $B_1 = 1$, $R_P = 1$, $\alpha = 6°$, $n_1 = N_{3D}/2$. $TR$=2500ms, $TE$=2.15ms, $TI$=100ms, $ES$=10.2 ms.

$$T_1^* = \left[\frac{1}{T_1} - \frac{1}{ES} \cdot \ln(\cos(\alpha \cdot B_1))\right]^{-1} \quad (4)$$

$$\tau_1 = TI - ES \cdot n_1 \quad (5)$$

$$TR = \tau_1 + ES \cdot N_3D + \tau_2 \quad (6)$$

$$\tau_2 = TR - \tau_1 - ES \cdot N_3D \quad (7)$$

$$E_1 = e^{\left(\frac{-\tau_1}{T_1}\right)}, \ E_2 = e^{\left(\frac{-\tau_2}{T_1}\right)}, E_3 = e^{\left(\frac{-(N_3D \cdot ES)}{T_1^*}\right)}, E_4 = e^{\left(-\frac{(n_1 \cdot ES)}{T_1^*}\right)} \quad (8)$$

$$Q = \frac{\left(E_4[1 - 2E_1 + E_1E_2] + \left(\frac{T_1^*}{T_1}\right)[1 + E_1E_2E_3 - E_1E_2E_4 - E_4]\right)}{(1 + E_1E_2E_3)} \quad (9)$$

$$S_{\text{Syn-P}\ /0} = M_0 \sin(\alpha B_1) R_P Q \quad (10)$$

For Syn-PD1(sMPR_PD_T1), $M_0 = PD$, and For Syn-PD0(sMPR_T1), $M_0 = 1$.

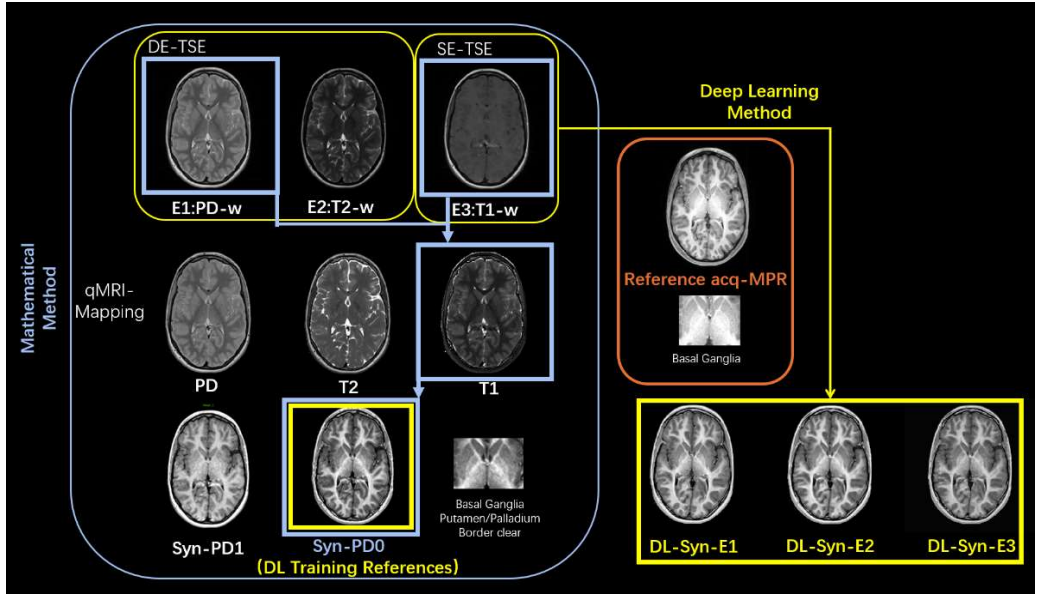

Fig.1. Workflow of Syn-PD0 contrast calculation, two steps necessary for Syn-PD0 calculation are squared in blue color. The basal ganglia regions of two contrasts are extracted for detail comparison. The necessary input, training reference, corresponding output are squared in yellow color. Syn-PD0 have a clearer border between putamen and palladium compared to the reference acquired MP-RAGE as marked in the orange box.

After truncating the low signal background, we normalized the data from 0 to 1, this measure diminishes the variation in input and label data, which is important for effective training[28].

*2.3. Network Implementation*

In this study, the core workflow (marked in Fig.2(e)) involves utilizing a conditional Generative Adversarial Network (cGAN/pix2pix) framework, to synthesize desired contrasts. For the training, the generator network (marked in Fig.2(a)) accepts a 4-channel different contrast (Tri-TSE, Syn-PD0) tensor, and Syn-PD0 were selected as the synthesis target and Tri-TSE were chosen as network input separately. The discriminator (marked in Fig.2(b)), meanwhile, receives both the real image pairs (marked as green, ground truth Syn-PD0 and input TSE images) and the generated image pairs (marked as orange, generated DL-Syn-MPR and input TSE images) to distinguish between real and fake as a binary classifier.

*2.4. Network Architecture Overview*

Our generator network builds upon the UNET architecture (marked in Fig.2(a)), known for its effectiveness in image-to-image translation, preserving both texture detail and perceptual quality. The network uses an encoder-decoder architecture that captures both global context and fine-grained details. To enhance its non-linear approximation capabilities, we have incorporated components inspired by Kolmogorov-Arnold Networks (KAN)[34], specifically by integrating Tokenization-KAN-Blocks (Tok-KAN), inspired by Li et. Al's U-KAN[35]. These modules are embedded in the bottleneck of the encoder-decoder pathways for enhancing.

The generator architecture is distinct from a conventional UNET in several key aspects as marked in Fig.2(d):

Patch Embedding Layer: After the initial three double convolutional neural network (CNN) layers in the encoding path, the network employs three levels of Patch Embedding, corresponding to embedded sizes of 128, 256, and 512, respectively. This layer divides the input image into two-dimensional patches of size $7 \times 7$ with a stride of 4 and projects these $7 \times 7$ patches into one-dimensional token sequences via a convolutional layer. The resulting tokens are flattened and subjected to Layer Normalization. Unlike the multi-head attention mechanism used in transformer architectures, this embedding approach leverages convolution for local feature extraction.

KAN-Blocks: Following patch embedding, the tokens are processed through KAN-Blocks, each consisting of two main components: Kolmogorov-Arnold Network (KAN) layers and Depth-wise Convolutions (DwConv). The KAN layers utilize B-splines for functional approximation, enabling efficient linear projection of complex, high-dimensional features into lower-dimensional representations while preserving accuracy. Each Tok-KAN block contains three KAN layers, interleaved with Depth-wise Convolutions and ReLU activations to improve feature representation. Residual connections are incorporated after each projection to stabilize the gradient flow by allowing gradients to bypass non-linear transformations, thus mitigating the vanishing gradient problem and facilitating efficient training of deeper networks.

Depth-wise Convolutions with BN-ReLU (DwC): Depth-wise Convolutions are utilized to efficiently capture local features while minimizing the number of parameters. Batch Normalization (BN) and ReLU activation follow each convolution to enhance the non-linearity of the feature representations.

Decoding Path: After processing through the KAN-Blocks, tokens enter the decoding path, which utilizes standard CNN upsampling layers like those in UNET to reconstruct the spatial dimensions and produce the final output image.

Compared to U-KAN, which employs tensor addition in skip connections, our KAN-UNET adopts tensor concatenation between the encoding and decoding paths. Tensor concatenation allows the network to retain more comprehensive feature information, which enhances the richness of representation and ultimately improves the quality of synthesis. This modification enhances the extraction of detailed features, thereby improving the quality of the synthesized images.

For the discriminator we implemented a traditional PatchGAN discriminator, which is commonly used in pix2pix models. The key architectural components of the discriminator consist of three CNN blocks, with LeakyRelu of slope of 2. The layers use a kernel size of 4 and a stride of 2. Then the output is matched with a sigmoid activation layer to decide whether the input pair is real or synthesized (fake).

## 2.5. Loss Functions

We utilized several loss functions to guide the training. For the part of the generator, firstly we have the L1 Loss measures the absolute differences between the predicted output ($y_{\text{fake}}$) and the ground truth ($y$), promoting pixel-wise accuracy:

$$L_{L1} = E_{x \sim p_{\text{data}}(x)} \lambda_{L1} \, \| y_{\text{fake}} - y \|_1 \tag{11}$$

where $\lambda_{L1} = 500$ is the weight assigned to the L1 Loss. $E$ is the maximum likelihood estimate.

BCE Loss guides the generator to produce outputs that can deceive the discriminator:

$$L_{G_{\text{fake}}} = E_{x \sim p_{\text{data}}(x)} \text{BCE}(D_{\text{fake}}, 1) \tag{12}$$

Here, $D_{\text{fake}}$ represents the discriminator's output for the generated data, and 1 is a tensor of ones.

To enhance edge information in the generated images, inspired by the work EaGAN[27], we employed the Sobel Edge Loss that:

$$L_{\text{Sobel}} = E_{x \sim p_{\text{data}}(x)} \lambda_{\text{edge}} \, \text{Sobel Loss}\left(\text{edge}_y, \text{edge}_{y_{\text{fake}}}\right) \tag{13}$$

The total loss for the generator combines all the individual losses:

$$L_{G,\text{total}} = L_{G_{\text{fake}}} + L_{L1} + L_{\text{Sobel}} \tag{14}$$

For the discriminator, its goal is to maximize the probability that the ground truth is judged to be a real label, while minimizing the probability that the generated fake sample is judged to be a real sample.

$$L_D = E_{x \sim p_{\text{data}}(x)}\big(\text{BCE}(D_{\text{real}}, 1)\big) + E_{x,y \sim p_{\text{data}}(x,y)}\big(\text{BCE}(D_{\text{fake}}, 0)\big) \tag{15}$$

$$L_{\text{total}} = L_D + L_{G_{\text{fake}}} + L_{L1} + L_{\text{Sobel}} \tag{16}$$

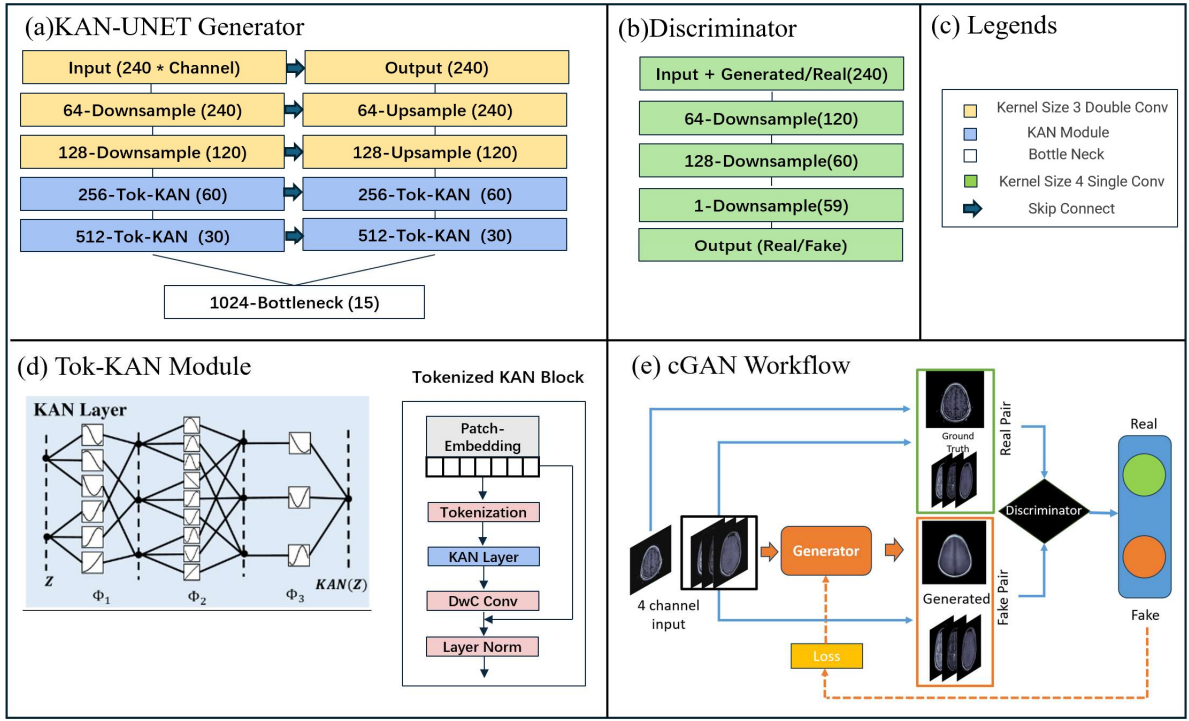

Fig.2. Workflow of the deep learning synthesis of Syn-MP-RAGE contrast. (a). KAN-UNET Generator, (b) Discriminator, (c) Legends of (a) and (b), (d) Tokenized-KAN Module and KAN Layer in KAN-UNET Generator, KAN Layer figure used the reference of U-KAN, (e) cGAN workflow overview

## 2.6. Training Detail

During training, we employed the Adam optimizer with a learning rate of $1 \times 10^{-4}$ and a batch size of 8. The L1 loss was weighed with $\lambda_{L1} = 500$, and the edge loss was weighted with $\lambda_{edge} = 100$. These hyperparameters were selected to balance accuracy and edge preservation in the generated images. the generator and discriminator are updated in alternating steps. First, the discriminator is trained to better distinguish between real and fake image pairs. Then, the generator is updated to improve its ability to produce convincing fake images, specifically by minimizing the discriminator's ability to classify them as fake, while also minimizing the loss. The entire training process leverages gradient scaling techniques to stabilize training.

All experiments were performed on an RTX 3090 GPU (24 GB memory), PyTorch version 1.12.1 and CUDA 11.7. For comparative analysis, we evaluated several models, including KAN-UNET, UNET, along with U-KAN and cycleGAN, some employed KAN-module as well and the other one commonly used in unsupervised image translation work. KAN-UNET, UNET, and U-KAN were implemented as generators in cGAN framework, cycleGAN was implemented as model. Each model was trained for 100 epochs. Data augmentation techniques, including random flipping, rotation, and cropping, were applied to improve the robustness of the training.

## 2.7. Evaluation Metrics

For the preliminary image quality evaluation, we included the metrics of Mean Average Error (MAE), Peak Signal-to-Noise Ratio (PSNR)and Structure Similarity Index Measure (SSIM), Paired student's t-test was performed between the metrics of other models and the proposed KAN-UNET model, the best result is bolded, and the significant result is marked with *. MAE reflects pixel-wise differences, and it's defined as:

$$\text{MAE} = \frac{1}{n}\sum_{i=1}^{n}|y_i - \hat{y}_i| \tag{17}$$

Here, $n$ represents the total number of pixels evaluated. The term $y_i$ is the actual pixel value of the i-th data point in the ground truth image, while $\hat{y}_i$ corresponds to the predicted pixel value in the synthesized image.

PSNR is used to measure the strength of error between the target image and the source, it is defined as:

$$\text{PSNR} = 10\log_{10}\left(\frac{\max I^2}{\text{MSE}}\right), \text{MSE} = \frac{1}{n}\sum_{i=1}^{n}(y_i - \hat{y}_i)^2 \tag{18}$$

Here $\max I^2$ denotes the max image intensity of the compared images. The higher the PSNR, the lower the noise of the image and the closer the result is to the reference image.

SSIM evaluates the visual similarity of two images, especially their structural information:

$$\text{SSIM} = \frac{(2\mu_1\mu_2 + C_1)(2\sigma_{12} + C_2)}{(\mu_1^2 + \mu_2^2 + C_1)(\sigma_1^2 + \sigma_2^2 + C_2)} \tag{19}$$

$\mu_1\mu_2$, $\sigma_1^2\sigma_2^2$ are the mean values and variance of the images 1 and 2, respectively, $\sigma_{12}$ is the covariance between the image 1 and 2, $C_1$ and $C_2$ are two constants preventing the denominator from being zero. SSIM evaluates the structural similarity of an image, taking into account intensity, contrast and structural information, which can better reflect the perception of image quality by the human eye. The higher the SSIM value, the closer the synthesized image is to the real image in perception.

We further performed FreeSurfer brain segmentation on Syn-PD0 and DL-Syn-MPR, calculated the corresponding overlap using Dice Similarity Coefficient (DSC). Both the subject group with metal-artifact and without metal artifact were calculated.

DSC was used to measure the similarity of segmentation overlap between the DL-Syn contrast and Syn-PD0 contrast. Its calculation formula is:

$$\text{DSC} = \frac{2|A \cap B|}{|A|+|B|} \tag{20}$$

Where $A$ is the pixel set of first segmentation result, $B$ is the pixel set of second segmentation result, $|A \cap B|$ indicates the number of pixels in the intersection of two segmentation results. The value of DSC ranges from 0 to 1, and the closer the value is to 1, the more similar the two segmentation results are, and higher the quality of the synthesis.

## 3. Results

### 3.1. Preliminary Synthesis

All the applied deep learning models produce the grey matter/ white matter microstructure originally unobservable in the input modalities of TSE, and synthetic image quality varies significantly across different generators, with KAN-UNET performing the best qualitatively in terms of detail retention, smoothness, and intensity control, demonstrating stable outputs and fewer distortions, particularly in complex anatomical structures retention like the basal ganglia as marked in Fig.4(d). Overall, E3 (T1-weighted inputs) is the most difficult to use, resulting in noisier outputs and higher discrepancies across models, whereas E1 (PD-weighted) and E2 (T2-weighted) inputs produce smoother outputs with fewer noise artifacts, but still using E3 input KAN-UNET model can remain the visibility of the boundary between the pallidum and putamen as marked by the red arrows. The hot spot difference maps reveal that generators like, U-KAN, cycleGAN and UNET introduce more background noise, blurs and discrepancies. while KAN-UNET and UNET retain these details more accurately. Sometimes high difference value observable in the boundary of the skull, but for the ROI of the brain inner structure the difference is lower.

Moreover, KAN-UNET outperforms UNET quantitatively in all input cases according to the metrics in Table 1 and Fig.3, suggesting that the Tok-KAN module may better leverage features and control the background intensity, particularly when handling both global and local information with the help of patch-embedding layer. While sometimes UNET showed a relative comparable result for specific metrics for E3 input, but its performances are not stable in E1, E2 metrices. It's possible that KAN Layer and Tokenization within the KAN-UNET architecture will likely allow for richer feature extraction, thereby improving synthesis quality. Overall KAN-UNET showed the most promising result in generating DL-Syn contrasts, so we employed it for further studies.

The KAN-UNET model required approximately 8 minutes per epoch during training, total training time for a single modality is 13 hours over 8160 slices. For inference, KAN-UNET took about 9 seconds to generate a prediction for a single modality that consists of 160 slices. The remaining 44 subjects were synthesized using the KAN-UNET for all three input contrasts.

Table 1 Evaluation metric of models to generate DL-Syn contrast with different inputs.

| Input | Metric | KAN-UNET | U-KAN | UNET | cycleGAN |
|---|---|---|---|---|---|
| E1 | MAE | 0.016 ± 0.006 | 0.048 ± 0.017 * | 0.027 ± 0.008 * | 0.028 ± 0.009 * |
| E1 | PSNR | 27.086 ± 2.245 | 19.530 ± 2.457 * | 23.190 ± 2.091 * | 23.234 ± 2.320 * |
| E1 | SSIM | 0.956 ± 0.019 | 0.867 ± 0.043 * | 0.924 ± 0.026 * | 0.922 ± 0.033 * |
| E2 | MAE | 0.011 ± 0.004 | 0.027 ± 0.008 * | 0.031 ± 0.011 * | 0.024 ± 0.007 * |
| E2 | PSNR | 29.985 ± 2.344 | 22.719 ± 1.706 * | 21.900 ± 2.606 * | 24.103 ± 1.769 * |
| E2 | SSIM | 0.971 ± 0.013 | 0.899 ± 0.032 * | 0.894 ± 0.041 * | 0.924 ± 0.030 * |
| E3 | MAE | 0.010 ± 0.004 | 0.026 ± 0.009 * | 0.012 ± 0.004 * | 0.033 ± 0.014 * |
| E3 | PSNR | 30.395 ± 2.409 | 23.728 ± 2.244 * | 29.771 ± 2.339 * | 22.099 ± 2.779 * |
| E3 | SSIM | 0.976 ± 0.011 | 0.905 ± 0.040 * | 0.975 ± 0.011 * | 0.893 ± 0.050 * |

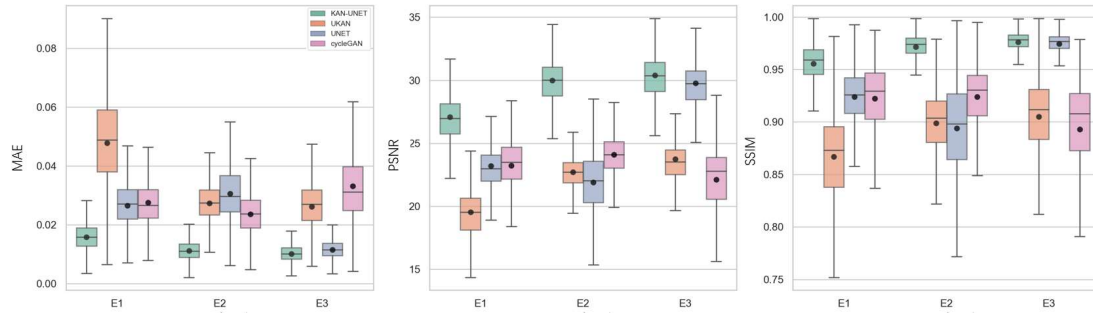

Fig.3. Metric boxplot of the three evaluation metrices.

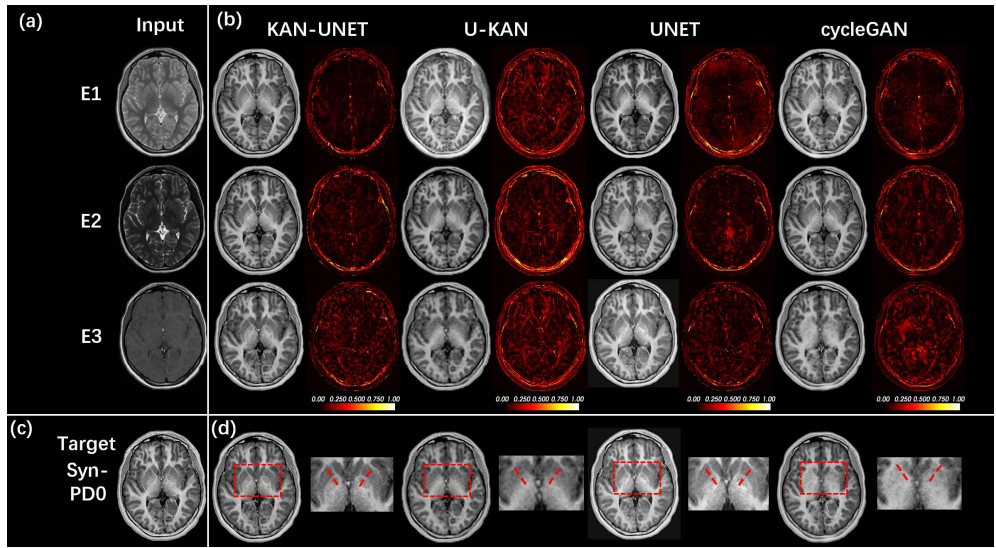

Fig.4. Visual Comparison of TSE image inputs (a), DL-Syn results and corresponding difference hot maps based on different generator (b). The first column and three rows show the input TSE modalities of E1(PD-weighted), E2(T2-weighted) and E3(T1-weighted). The last row of the first column (c) shows the target of DL training. The basal ganglia region of DL-Syn-MPR by E3 input were squared in red dashed lines, and corresponding area of basal ganglia emphasis were juxtaposed right next in (d).

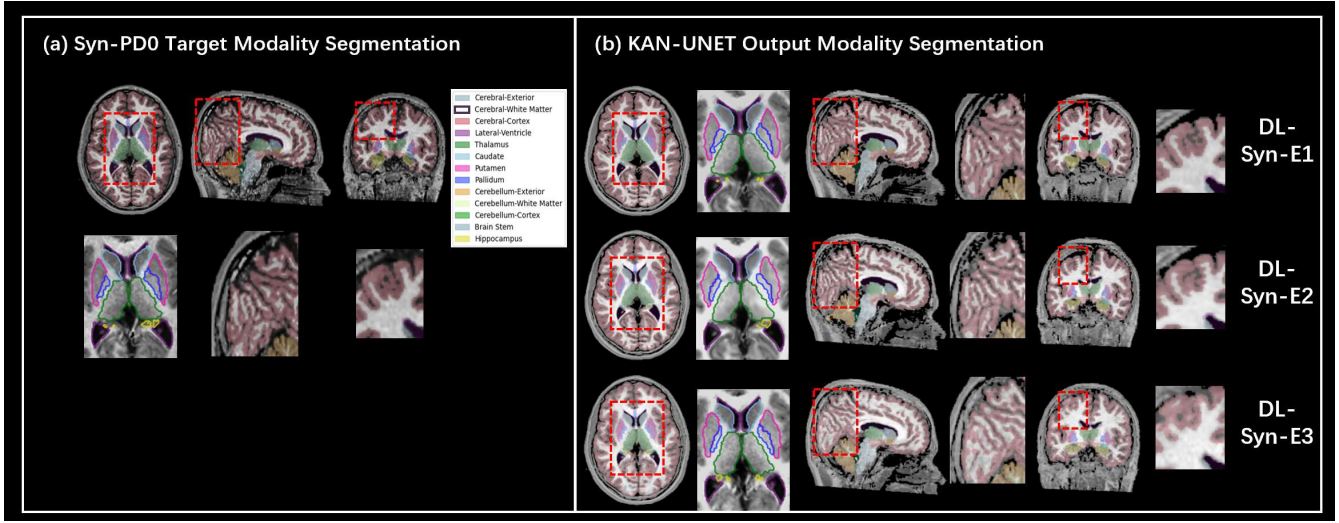

Fig.5. Triple orthogonal segmentation map overlay of the DL-Syn-MPR modalities with FreeSurfer colour legend of an example subject. The basal ganglia in the axial view, occipital region in the sagittal view and the precentral gyrus of the coronal view were squared in the red dashed line for detailed analysis. Training reference/target Syn-PD0 shown in (a). corresponding DL-Syn-MPR output contrast shown in (b).

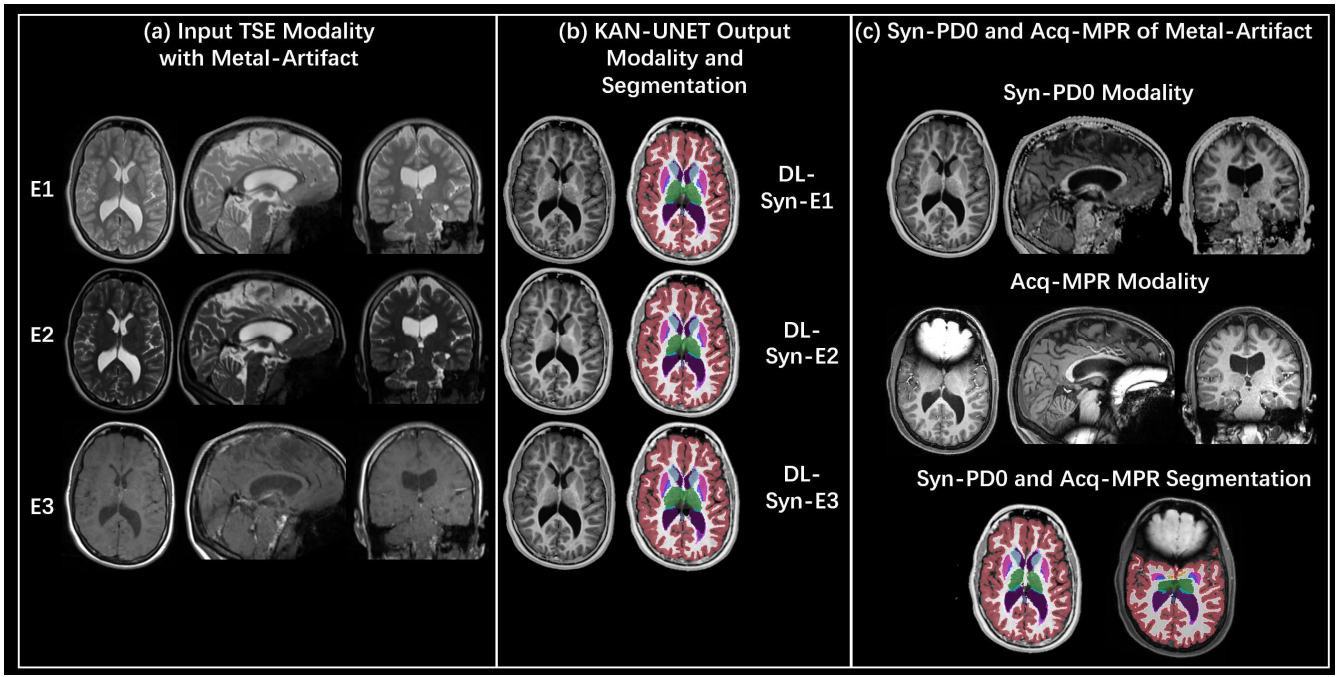

Fig.6. Metal-artifact subject overview. Input Tri-TSE modalities triple orthogonal view are shown in (a), DL-Syn-MPR modalities' axial view done by KAN-UNET shown in (b), the training reference Syn-PD0 and Acquired MP-RAGE contrast shown in (c).

### 3.2. Freesurfer Segmentation

To further assess the potential of the synthesized contrast, cerebral segmentation was performed on both Syn-PD0 and DL-Syn contrasts. The segmentation process was conducted using FreeSurfer version 7.4.1 on an Ubuntu 22.04 system, with each subject's segmentation taking approximately 2 hours. All three DL-Syn contrasts successfully performed the brain segmentation. From Fig.5 (b), we can see that DL-Syn-E1 and E2 segmentation better preserve the similarity to Syn-PD0 in Fig.5 (a).

However, DL-Syn-E3 shows issues such as ballooning of gyri in the sagittal view and discontinuity in the coronal view. The morphological shape of the pallidum and putamen also differs from Syn-PD0, indicating insufficient contrast for segmentation.

Segmentation was also performed on subjects with metal artifacts as shown in Fig.6. Compared to the acquired MP-RAGE, TSE sequences, which are less affected by metal implants with the advantage of spin-echo sequence, also result in DL-Syn and Syn-PD0 contrasts that are less influenced by artifacts as shown in Fig.6(a). Thus, these synthesized contrasts produce more accurate segmentations compared to the acquired MP-RAGE, where significant signal loss in the frontal sinus region can be observed in acquire MP-RAGE when juxtaposing Fig.6(b) and Fig.6(c).

The Dice Similarity Coefficient (DSC) was calculated for 42 subjects (31 without artifacts and 11 with metal artifacts) from another medical centre. We selected the 7 most representative brain regions in the basal ganglia for calculation: White Matter (WM), Gray Matter/Cortex (GM), Ventricle (VE), Caudate Nucleus (CN), Thalamus (Th), Putamen (Pu), and Pallidum (Pa). The quantitative results shown in the boxplot of Fig.7 indicate that E1 consistently achieves the highest DSC values compared to the other two input contrasts, with most anatomical regions having DSC scores above 0.83. The Pa scored lowest for all three inputs, likely due to its smaller volume compared to WM/GM, and also possibly due to the non-isotropic acquisition affecting continuity along the z-axis. E2 ranks in the middle in terms of performance, while E3 shows the lowest DSC values, which contrasts with earlier findings from the quantitative image quality comparison.

This apparent contradiction could be partly attributed to the noise transfer mechanism, specifically the propagation of error input modalities and the ground truth during synthesis as shown in equation (21) $\sigma$ represent noise in this case[6]. In particular, the noise pattern in E3 input appears more prominent in the ground truth Syn-PD0 modality, leading to higher quantitative scores for E3. However, this higher quantitative score did not necessarily reflect better segmentation performance, as the noise in E3 does not carry useful edge or anatomical contrast, unlike E1.

$$\sigma_{\text{Syn-P}} = \sqrt{((\partial y/\partial x_{E_1})^2 \sigma_{E_1}^2 + (\partial y/\partial x_{E_3})^2 \sigma_{E_3}^2)} \qquad (21)$$

For the acquired MP-RAGE images, metal artifacts cause severe signal loss and distortion in the prefrontal lobe area of the brain. Rapid changes in the static magnetic field induce signal dephasing, resulting in regions of signal loss or distortion in the images[36], [37], which severely affects the segmentation accuracy as shown in Fig.6(c). In such case Syn-PD0 contrasts are more useful compared to acquire-MP-RAGE.

In contrast, across the various synthetic modalities, although input images are still partially affected by artifacts (low signal black area), the TSE inputs' relatively strong resistance to artifacts ensures better preservation of anatomical structures. In both the synthesized (Syn-PD0) and KAN-UNET output DL-Syn results, the prefrontal lobe remains well-preserved compared to the acquired counterparts, with clear anatomical delineation in the basal ganglia region. For the DSC score, no significant differences were found between the metal-artifact group and the non-artifact group as indicated in Table 2 and Fig 7.

## 4. Discussion

The results from this study indicate that the KAN-UNET model successfully synthesizes high-quality synthetic MP-RAGE (DL-Syn-MPR) images from Tri-TSE inputs. Leveraging the advanced fitting capabilities of the KAN modules, particularly in managing complex anatomical structures, the model demonstrates significant improvements over traditional (like UNET generator and cycleGAN) or similar deep learning framework (like U-KAN) in image synthesis quality. Quantitatively, KAN-UNET achieved superior results in terms of MAE, PSNR, and SSIM across all input modalities (E1, E2, and E3). The qualitative analysis shows that KAN-UNET we proposed here consistently retains structural details in regions of interest, especially in the basal ganglia, to a better extent compared to other models. Comparison of DL-Syn and the Syn-PD0's segmentation reveals that even though the DL-Syn-E3 showed higher evaluation metric, DL-Syn-E1 and E2 better preserve the fidelity of the Syn-PD0 morphologically. And DL-Syn-E1 showed a high DSC score in the range from 0.83 to 0.93 in the ROI of basal ganglia. The result in the artifact comparison group suggests that the DL-Syn-MPR also preserves the metal-resistant ability of Syn-PD0 with no significant difference in DSC score.

The input flexibility brought by the DL-Syn-MPR could be vital in the clinical scenario. The use of the previously underutilized TSE-E2 modality in generating DL-Syn-E2 provides a flexible option for enhanced synthetic image generation. The foremention input flexibility of DL-Syn contrast allows for handling corrupted input cases effectively. As shown in Fig.8(a)'s example, a single TSE input image exhibits motion artifacts, resulting in a ring-like texture in the Syn-PD0 output, whereas the DL-Syn contrasts (E1/E2) remain unaffected due to the independent input. Similarly, in Fig.8(b), a misalignment of the TSE input of 11.23 mm vertically results in ghosting effects in the corresponding Syn-PD0 calculation, while the DL-Syn (E1, E2, and E3) images shown unaffected. Based on the dataset of ELGAN-ECHO, as shown in Fig.8(c), among 401 subjects, 10 subjects are misaligned in registration, 8 subjects have motion artifacts. These datasets with artifacts could be used again with the help of our method.

Table 2. DSC value of seven representative basal ganglia regions with different input experiments, for metal-artifact subject group and non-subject group.

| ROI | E1 | | E2 | | E3 | |
|---|---|---|---|---|---|---|
| | Non-Artifact | Artifact | Non-Artifact | Artifact | Non-Artifact | Artifact |
| White-Matter | 0.934 ± 0.02 | 0.92 ± 0.03 | 0.91 ± 0.02 | 0.891 ± 0.033 | 0.868 ± 0.05 | 0.814 ± 0.067 |
| Cortex | 0.879 ± 0.03 | 0.844 ± 0.05 | 0.863 ± 0.04 | 0.822 ± 0.057 | 0.818 ± 0.06 | 0.746 ± 0.068 |
| Ventricle | 0.926 ± 0.02 | 0.896 ± 0.05 | 0.922 ± 0.03 | 0.89 ± 0.046 | 0.892 ± 0.05 | 0.825 ± 0.093 |
| Thalamus | 0.919 ± 0.01 | 0.909 ± 0.02 | 0.899 ± 0.01 | 0.888 ± 0.021 | 0.879 ± 0.02 | 0.856 ± 0.06 |
| Putamen | 0.877 ± 0.07 | 0.905 ± 0.03 | 0.85 ± 0.05 | 0.87 ± 0.024 | 0.834 ± 0.06 | 0.796 ± 0.103 |
| Pallidum | 0.824 ± 0.06 | 0.844 ± 0.04 | 0.75 ± 0.06 | 0.77 ± 0.036 | 0.749 ± 0.07 | 0.705 ± 0.119 |
| Caudate | 0.913 ± 0.01 | 0.893 ± 0.04 | 0.888 ± 0.02 | 0.864 ± 0.048 | 0.852 ± 0.04 | 0.794 ± 0.108 |

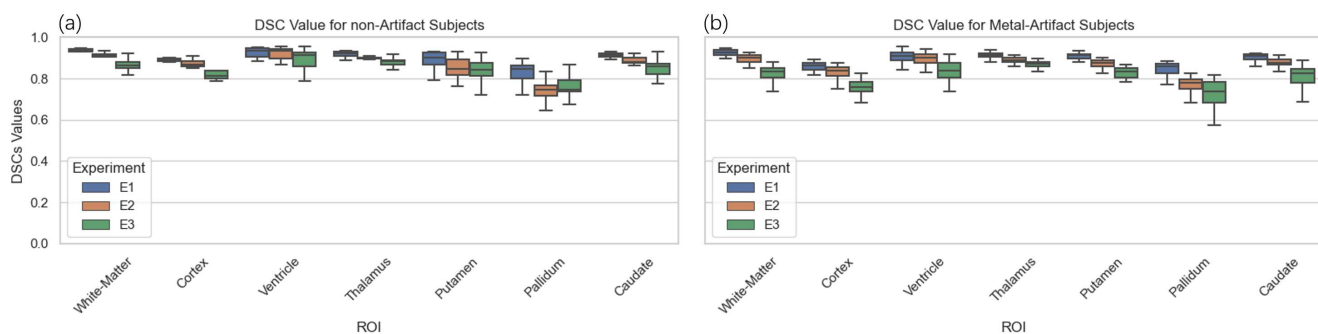

Fig.7. DSC value of seven representative basal ganglia regions with different input experiments, the non-artifact subject shown in (a), metal-artifact subject shown in (b).

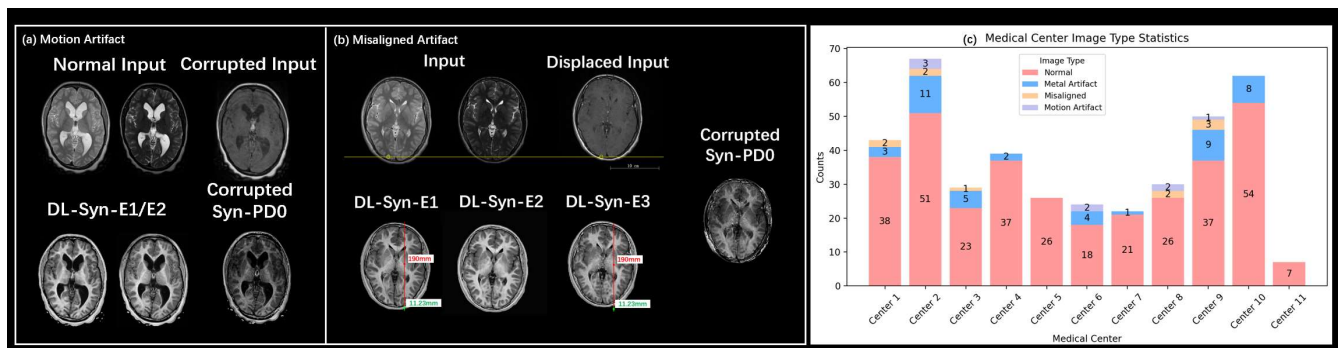

Fig.8. **Corrupted Input Cases.** (a) Motion artifact: Single TSE input image exhibits motion artifacts. (b) Misaligned artifact: The DE-TSE input is vertically misaligned by 11.23 mm with SE-TSE input.

In summary, the use of DL-Syn-MPR offers unique advantages, such as reducing the need for multiple scans and complex registration processes when generating metal-artifact resistant Syn-PD0 with single channel input. However, there are certain limitations to the study:

2D UNET Backbone: This study employed 2D UNET architecture, which, while effective and computationally efficient, may miss important spatial information in the sagittal and coronal directions. This limitation could be a factor in the relatively low DSC scores observed in small-volume brain regions, such as the pallidum. A 3D UNET could better capture volumetric relationships and improve synthesis quality.

Interpolation for Isotropic Resampling: To achieve isotropic resolution, the Tri-TSE images, which were acquired in 2D with non-isotropic voxel size, were interpolated to $1mm^3$ resolution. This interpolation may have introduced blurring effects in sagittal and coronal views. Using fully isotropic Tri-TSE images for future training could mitigate this issue.

Unified Training for Input Modalities: Currently, each modality was trained separately. Future work could explore methods to simplify training so that a single model can be used for all input predictions, improving efficiency.

Larger Subject Cohort: This study was conducted on a limited subject cohort. Future studies with larger-scale datasets provide more robust insights into model generalization and performance across diverse populations.

Pathological Studies: While the generated synthetic contrasts have shown promising results in normal anatomical regions, further clinical validation is needed for pathological conditions, such as epilepsy or multiple sclerosis, where precise volumetric accuracy is critical.